# Model uncertainty in measuring the gluon jet fraction at the hadron collider

S. G. Shulha (https://orcid.org/0000-0002-4265-928X)

*Joint Institute for Nuclear Research, Dubna, 141980 Russia*

*F.Skorina Gomel State University, Gomel, 246019 Belarus*

*e-mail: shulga@jinr.ru*

**Abstract**— The work proposes a new method for measuring the gluon jet fraction in jet sample produced at the hadron collider. This method uses model quark/gluon templates - distributions of quark and gluon jets over the jet macro parameter. Within the framework of this method, it is possible to find a model uncertainty of measurement associated with the deviation of model quark/gluon templates from the true ones. The issue of data-motivated corrections to the model quark/gluon templates is discussed.

## INTRODUCTION

This work continues a series of articles [1-3] dedicated to the methodology for measuring the gluon jet fraction in a sample of jets produced at the hadron collider. Paper [1] discusses a method that uses fitting of the measured distribution of jets over $q/g$-discriminator using linear combination of model quark ($q$) and gluon ($g$) jet distributions ($q/g$-templates). To measure the fraction, one can use any macro parameters (MPs) of the jet instead of the $q/g$-discriminator. However, modeling shows [3] that the fitting method gives a stable result for $q/g$-discriminator, but not for simple jet MPs. This paper proposes a method for measuring $g$-fractions that does not use fitting. This method allows you to use any jet MPs to measure $g$-fraction. If one measures $g$-fraction with different MPs, then the variations in the resulting $g$-fractions can be considered as the model uncertainty in the measurement of the $g$-fraction.

The article contains three sections. The first section is dedicated to a description of new method for measuring $g$-fraction. Model uncertainty of $g$-fraction measurement is discussed in the second section. The third section discusses the issue of $q/g$-template corrections, which is an important part of the work on $q/g$-discriminators at the CMS [4, 5].

## METHOD OF "AVERAGING" IN MEASURING THE FRACTION OF GLUON JETS

Quark and gluon jets have significantly different macro parameters (MP), $V$, and any jet sample can be considered as a combination of subsamples of $q$-jets and $g$-jets. Distribution of jets over $V$ has the form of a bin-by-bin sum of $q/g$-subsample histograms:

$$h_{\text{MC}}(V) = h^g_{\text{MC}}(V) + h^q_{\text{MC}}(V), \tag{1}$$

where the number of entries in $h_{\text{MC}}$, $N$, is equal to the sum of the numbers of entries in $h^g_{\text{MC}}$ and $h^q_{\text{MC}}$: $N = N^g + N^q$. Index "MC" stands for Monte Carlo jets ("MC jets") which are produced by MC generator, passed through the simulation of the detector response and then reconstructed. Analysis at the level of reconstructed jets (instead of generator ones) allows one to avoid the "unfolding" procedure for jet distributions in data, $h_{\text{DAT}}(V)$.

The flavor of the MC jet is determined by the geometric matching of the most energetic parton and the MC jet. Here it is assumed that jets with an unidentified flavour ($x$-jets) included in subsamples $q$-jets or $g$-jets. Jet flavour identification and measuring the $x$-jet fraction are discussed in detail in [2].

The normalized distribution of MC jets, $H_{\text{MC}}$, has the form:

$$H_{\text{MC}}(V) = \alpha^g_{\text{MC}} H^g_{\text{MC}}(V) + \left(1 - \alpha^g_{\text{MC}}\right) H^q_{\text{MC}}(V), \tag{2}$$

where $\alpha^g_{\text{MC}} = N^g/N$ is $g$-jet fraction, $(1 - \alpha^g_{\text{MC}})$ - $q$-jet fraction. Normalized distributions of $q/g$-jets, $H^{q/g}_{\text{MC}}$, will also be called $q/g$-templates for brevity.

For real jets reconstructed in detector ("DAT jets"), an equation like (2) has the form:

$$H_{\text{DAT}}(V) = \alpha^g_{\text{DAT}} H^g_{\text{DAT}}(V) + \left(1 - \alpha^g_{\text{DAT}}\right) H^q_{\text{DAT}}(V), \tag{3}$$

On the right side of Eq. (3) there are three unknown "quantities": $\alpha^g_{\text{DAT}}$ and two $q/g$-templates. If we consider $K$ jet samples with identical $q/g$-templates and different $\alpha^g_{1,\dots,K;\text{DAT}}$ and write down a system of $K$ equations similar to Eq. (3), then the number of unknowns is equal to $K + 2$. It is impossible to extract (measure) both $q/g$-templates and $g$-jet fractions from the data.

This is not accidental, as an additional assumption is required about the shape of the objects ($q/g$-jets) that are recognized in the data. To determine the $g$-jet fractions in the data, $\alpha^g_{\text{DAT}}$, you are required to use MC $q/g$-templates:

$$H_{\text{DAT}}(V) = \alpha^g_{\text{DAT}} H^g_{\text{MC}}(V) + \left(1 - \alpha^g_{\text{DAT}}\right) H^q_{\text{MC}}(V). \tag{4}$$

However, exact equality in Eq. (4) is possible for $\alpha^g_{\text{DAT}}$ that depends on the $V$-bin: $\alpha^g_{\text{DAT}}(V)$. This dependence can be interpreted as statistical and systematic uncertainty in the measurement of $\alpha^g_{\text{DAT}}$, and each $V$-bin represents an "independent experiment" in this measurement. From Eq. (4) we find:

$$\alpha^g_{\text{DAT}}(V) = \frac{H_{\text{DAT}}(V) - H^q_{\text{MC}}(V)}{H^g_{\text{MC}}(V) - H^q_{\text{MC}}(V)}. \tag{5}$$

Measured $g$-jet fraction is averaged value with standard error:

$$\alpha^g_{\text{DAT}} = \langle \alpha^g_{\text{DAT}}(V) \rangle, \quad \Delta\alpha^g_{\text{DAT}} = \frac{\sigma}{\sqrt{N_V}}, \quad \sigma = \sqrt{\langle \alpha^g_{\text{DAT}}(V)^2 \rangle - \langle \alpha^g_{\text{DAT}}(V) \rangle^2}, \tag{6}$$

where $N_V$ – number of $V$-bins used. Larger denominator in Eq. (5) results in smaller error, which is propagated from two sources: statistical deviations of measured $H_{\text{DAT}}(V)$ and systematic deviations of MC $q/g$-templates, $H^{q,g}_{\text{MC}}(V)$, from real unknown data $q/g$-templates.

In [1], it was proposed to use one-parametric fitting ($\sim$) in order to find $\alpha^g_{\text{DAT}}$:

$$H_{\text{DAT}}(V) \sim \alpha^g_{\text{DAT}} H^g_{\text{MC}}(V) + \left(1 - \alpha^g_{\text{DAT}}\right) H^q_{\text{MC}}(V), \tag{7}$$

Larger difference between MC $q$- and $g$-templates results in smaller uncertainty of fitting (7). A stable fit with relatively small statistics is achieved (see [3]) for combined jet MP "Quark-Gluon Likelihood" ($V = QGL$) defined in [3, 4]. Simulation shows that fit (7) for a simple jet MP, (e.g., particle multiplicity inside jet, jet angular size etc.) is unstable and requires significantly more statistics. The method described above and represented by Eqs. (5)-(6) is more stable. The acceptable difference between the templates in the denominator of Eq. (5) can be controlled in the numerical calculation so that the uncertainties do not exceed the acceptable limit. This method can be used for simple jet MP also.

The simplicity of method Eqs. (5)-(6) is another advantage of this method. There is no need to build a discriminator $QGL$ to measure $g$-fractions. On the other hand, measuring $g$-fractions using simple jet MPs can play the role of a test for the constructed $QGL$: the value of $\alpha_{DAT}^g$ obtained using $V = QGL$ should be in area of the $\alpha_{DAT}^g$ value averaged for all jet MPs that are used to construct $QGL$.

## MODEL UNCERTAINTY IN MEASUREMENT OF GLUON JET FRACTION

Measurement of g-fraction (Eq. (5) or (7)) is fundamentally model-dependent, since the $q/g$-templates are built using the model. Is it possible to quantify the model uncertainty for the measured g-fraction? The answer to this question is positive. The possibility of estimating the model uncertainty in the g-fraction measurement opens up the fact that various jet MPs can be used in Eq.(5).

The uncertainty of measuring the $g$-fraction Eq. (6) contains a standard statistical error and systematic uncertainty associated with the deviation of MC $q/g$-templates from true ones. If averaging in Eq. (6) extends over the bins of several MPs, then the systematic part of uncertainty $\Delta\alpha_{DAT}^g$ contains information about the deviation of all used MC $q/g$-templates (for all used MPs $V$) from the true $q/g$-templates. The systematic part of uncertainty $\Delta\alpha_{DAT}^g$ Eq. (6) obtained with the "full set of jet MPs" can be called model uncertainty. By definition, a set of MPs is full if adding any new MP does not increase uncertainty $\Delta\alpha_{DAT}^g$. Simulations show that the model uncertainty is several times larger than the uncertainty obtained with a single MP. For practical analysis, the MPs that are used to construct $QGL$ can be used as a "full set of jet MPs": the total multiplicity of particles inside jet, minor axis of the jet ellipse in 2-dimensional angular space, and MP $p_T D$ [4, 5].

## IS A DATA-MOTIVATED CORRECTION POSSIBLE FOR $q/g$-TEMPLATES?

If the fractions of $g$-jets are measured, then two unknown values remain in the Eq. (3): $q/g$-templates $H_{DAT}^q$ and $H_{DAT}^g$. To determine the $q/g$-templates, one can use two jet samples that are kinematically close to each other, but have different $g$-fractions. The kinematic similarity of the jet samples ensures the identity of the $q/g$-templates in the two jet samples. Two equations for two jet samples represent a system of equations for calculating $H_{DAT}^q$ and $H_{DAT}^g$:

$$H_{1,DAT}(V) = \alpha_{1,DAT}^g H_{DAT}^g(V) + (1 - \alpha_{1,DAT}^g) H_{DAT}^q(V),$$
$$H_{2,DAT}(V) = \alpha_{2,DAT}^g H_{DAT}^g(V) + (1 - \alpha_{2,DAT}^g) H_{DAT}^q(V). \quad (8)$$

The solution to the system of equations (8) has the form:

$$H_{DAT}^q(V) = \frac{\alpha_{2,DAT}^g H_{1,DAT}(V) - \alpha_{1,DAT}^g H_{2,DAT}(V)}{\alpha_{2,DAT}^g - \alpha_{1,DAT}^g}. \quad (9)$$

To write down $H_{DAT}^g$ one needs to make substitutions $\{g \to q, 1 \leftrightarrow 2\}$ in $H_{DAT}^q$.

In the current CMS analysis, that uses $q/g$-jet recognition, an important place is occupied by data-motivated corrections (so called "data-driven scale factors", SF) for MC $q/g$-templates [4, 5]:

$$S^f(V) \equiv \frac{H_{\text{DAT}}^f(V)}{H_{\text{MC}}^f(V)}. \qquad (10)$$

To obtain $S^f(V = QGL)$, a two-sample method based on Eqs. (8) is used in which $\alpha_{\text{MC}}^g$ is used instead of the measured $\alpha_{\text{DAT}}^g$ [4, 5]. Very large SFs for $g$-jets were found in [5]: $S^f(QGL) \in [0.75, 2.0]$. It is shown below that when using $\alpha_{\text{DAT}}^g$ in Eqs. (8)-(9), then $S^f(V) \approx 1$ within the uncertainty of measuring the $g$-fractions $\alpha_{1,\text{DAT}}^g$ and $\alpha_{2,\text{DAT}}^g$.

Indeed, let us write down two equations for one jet sample ($k = 1$ or $k = 2$):

$$H_{k,\text{DAT}}(V) \approx \alpha_{k,\text{DAT}}^g H_{\text{MC}}^g(V) + (1 - \alpha_{k,\text{DAT}}^g) H_{\text{MC}}^q(V)$$
$$H_{k,\text{DAT}}(V) = \alpha_{k,\text{DAT}}^g H_{\text{DAT}}^g(V) + (1 - \alpha_{k,\text{DAT}}^g) H_{\text{DAT}}^q(V). \qquad (11)$$

The left and right sides of the first equation (11) coincide within the measurement uncertainty of $\alpha_{k,\text{DAT}}^g$. The second equation is an identity, according to Eq. (8). Subtracting the first equation of system (11) from the second one, we obtain ($\Delta^f \equiv H_{\text{DAT}}^f - H_{\text{MC}}^f$):

$$\alpha_{k,\text{DAT}}^g \Delta^g + (1 - \alpha_{k,\text{DAT}}^g) \Delta^q \approx 0,$$

or

$$\alpha_{k,\text{DAT}}^g (\Delta^g - \Delta^q) \approx \Delta^q. \qquad (12)$$

Eq. (12) is satisfied for any $\alpha_{k,\text{DAT}}^g$ only if $\Delta^f \approx 0$. This is equivalent to $S^f(V) \approx 1$.

This results in two conclusions. First, there is no problem in obtaining the correction factor $S^f(V)$ using the data. Secondly, large values of $S^f(QGL)$ found in [5] indirectly indicate that the $g$-fractions in the data and in MC are very different. Using the values of $S^f(QGL) \in [0.75, 2.0]$ we can predict that $g$-fractions in the data are significantly smaller than in the model used in [4]. Preliminary results of analysis for Run-1 and Run-2 CMS data confirm this assumption.

## CONCLUSIONS

The paper proposes a new method for measuring the gluon jet fractions in jet sample, based on model $q/g$-templates, which is $q/g$-distributions over some jet macro parameter. Within the framework of this method, it is possible to determine the model uncertainty of measured $g$-fraction. The model uncertainty gives the lower bound of the theoretical uncertainty of measurements. It is shown, that using the measured $g$-fractions to obtain data motivated correction factor to model $q/g$-templates, gives 1 within the uncertainty of measuring the $g$-fractions. The significant deviations of the correction factor from 1, obtained earlier in the CMS collaboration, indirectly indicate that the generator $g$-fractions used in the calculation are significantly larger than data $g$-fractions.

## FUNDING

## CONFLICT OF INTEREST

The authors declare that they have no conflicts of interest.

`